\documentclass[letterpaper, 10 pt, conference]{ieeeconf}  
\IEEEoverridecommandlockouts                              

\usepackage{epsfig} 

\title{\LARGE \bf
Fourier Synthesis Methods for Control of Inhomogeneous Quantum
Systems }

\author{Brent Pryor and Navin Khaneja
\thanks{B. Pryor and N. Khaneja are with The School of Engineering and Applied Sciences,
        Harvard University
        {\tt\small pryor@fas.harvard.edu, navin@eecs.harvard.edu}}%
}

\begin{document}

\maketitle
\thispagestyle{empty}
\pagestyle{empty}

\begin{abstract}
Finding control laws (pulse sequences) that can compensate for
dispersions in parameters which govern the evolution of a quantum
system is an important problem in the fields of coherent
spectroscopy, imaging, and quantum information processing.  The use
of composite pulse techniques for such tasks has a long and widely
known history. In this paper, we introduce the method of Fourier
synthesis control law design for compensating dispersions in quantum
system dynamics. We focus on system models arising in NMR
spectroscopy and NMR imaging applications.
\end{abstract}

\section{Introduction}
Many applications in the control of quantum systems involve
controlling a large ensemble using the same control signal
\cite{Li_Khaneja, Bernstein}.  In many practical cases, the elements
of the ensemble show dispersions or variations in the parameters
which govern the dynamics of each individual system.  For example,
in magnetic resonance experiments, the spins in an ensemble may have
large dispersions in their resonance frequencies (Larmor dispersion)
or in the strength of the applied radio frequency fields (rf
inhomogeneity) seen by each member of the ensemble.  Another example
is in the field of NMR imaging, where a dispersion is intentionally
introduced in the form of a linear gradient \cite{Bernstein}, and
then exploited to successfully image the material under study.

A canonical problem in the control of quantum ensembles is the
design of rf fields (control laws) which can simultaneously steer a
continuum of systems, characterized by the variation in the internal
parameters governing the systems, from a given initial distribution
to a desired final distribution.  Such control laws are called
compensating pulse sequences in the Nuclear Magnetic Resonance (NMR)
literature.  From the standpoint of mathematical control theory, the
challenge is to simultaneously steer a continuum of systems between
points of interest using the same control signal.  Typical designs
include excitation and inversion pulses in NMR spectroscopy and
slice selective pulses in NMR imaging \cite{Bernstein, Levitt,
Tycko, Tycko2, Shaka, Levitt2, Levitt3, Garwood, Skinner, Kobzar,
Kobzar2}. In many cases, one desires to find a control law that
prepares the final state as some desired function of the parameters.
A premier example is the design of slice selective pulse sequences
in magnetic resonance imaging applications, where spins are excited
or inverted depending upon their physical position in the sample
under study \cite{Bernstein, SLR, Silver, Shinnar, LeRoux, Conolly}.
In fact, the design of such pulses is a fundamental requisite for
almost all magnetic resonance imaging techniques.  This paper
introduces the new method of Fourier synthesis pulse sequence design
for systems showing dispersions in the parameters governing their
dynamics.

In this paper we focus on the Bloch equations with a linear one
dimensional gradient, which arise in the context of NMR spectroscopy
and NMR imaging applications.
\begin{equation}
\frac{d}{dt}\left[\begin{array}{c} M_x  \\
M_y \\
M_z \end{array}\right] = \left[\begin{array}{ccc} 0 & -G(t) s & \epsilon u(t)  \\
G(t) s & 0 & -\epsilon v(t) \\
-\epsilon u(t) & \epsilon v(t) & 0 \end{array}\right]\left[\begin{array}{c} M_x \\
M_y \\
M_z \end{array}\right] \label{chap32_bloch}
\end{equation}
Here, $M(s, \epsilon) = [M_x \ M_y \ M_z]^T$ is the state vector,
$u(t) \in \Re$, $v(t) \in \Re$, and $G(t) \in \Re$ are controls and
the parameters $s \in [0, 1]$ and $\epsilon \in [1-\delta, 1], \
\delta > 0$ are dispersion parameters which will be explained
subsequently. Without loss of generality, we will always normalize
the initial state of the system (\ref{chap32_bloch}) to have unit
norm, so that the system evolves on the unit sphere in three
dimensions (Bloch sphere). A useful way to think about the Bloch
equations (\ref{chap32_bloch}), is by imagining a two dimensional
mesh of systems, each with a particular value of the pair
$(s,\epsilon)$. We are permitted to apply a single set of controls
$(u(t),v(t),G(t))$ to the entire mesh of systems, and the controls
should prepare the final state of each system as a desired function
of the parameters $(s,\epsilon)$ which govern the system dynamics.

From a physics perspective, the system (\ref{chap32_bloch})
corresponds to an ensemble of noninteracting spin-$\frac{1}{2}$ in a
static magnetic field $B_0$ along the $z$ axis and a transverse rf
field $(A(t) \cos(\psi(t)),A(t) \sin(\psi(t)))$ in the $x$-$y$
plane. The state vector $[M_x \ M_y \ M_z]^T$ represents the
coordinate of the unit vector in the direction of the net
magnetization vector for the ensemble \cite{Cavanagh}. The controls
$u(t)$ and $v(t)$ correspond to available rf fields we may apply to
the ensemble of spins. The dispersion in the magnitude of the rf
field applied to the sample is modeled by including a dispersion
parameter $\epsilon$ such that $A(t) = \epsilon A_0(t)$ with
$\epsilon \in [1-\delta, 1], \ \delta
>0$.  Thus, the maximum amplitude for the rf field ($\epsilon = 1$) corresponds
to the maximum amplitude seen by any spin in the ensemble.
Similarly, we consider a linear gradient $G(t) s$, where $G(t)$ may
be thought of as a control, and $s$ represents the normalized
spatial position of the spin system in the sample of interest.  In
(\ref{chap32_bloch}), we work in units with the gyromagnetic ratio
of the spins $\gamma = 1$.  In this paper we give new design methods
which scale polynomially that can be used to design pulse sequences
for (\ref{chap32_bloch}) which prepare the final state of the system
as a function of the parameters $s$ and $\epsilon$.

\section{Design Method for rf Inhomogeneity}
Considering only the Bloch equations with rf inhomogeneity and no
linear gradient ($G(t) = 0$), we can rewrite (\ref{chap32_bloch}) in
terms of the generators of rotation in three dimensions as
\begin{equation}
\frac{d}{dt}\left[\begin{array}{c} M_x  \\
M_y \\
M_z \end{array}\right] = \epsilon (u(t) \Omega_y + v(t) \Omega_x)  \left[\begin{array}{c} M_x \\
M_y \\
M_z \end{array}\right] \label{bloch_eps}
\end{equation}
where
\begin{eqnarray}
\Omega_x &=& \left[\begin{array}{ccc} 0 & 0 & 0  \\
0 & 0 & -1 \\
0 & 1 & 0 \end{array}\right] \nonumber \\
\Omega_y &=& \left[\begin{array}{ccc} 0 & 0 & 1  \\
0 & 0 & 0 \\
-1 & 0 & 0 \end{array}\right] \nonumber \\
\Omega_z &=& \left[\begin{array}{ccc} 0 & -1 & 0  \\
1 & 0 & 0 \\
0 & 0 & 0 \end{array}\right]
\end{eqnarray}
We will come back to the full version of the Bloch equations
(\ref{chap32_bloch}) with both a linear gradient and rf
inhomogeneity later in the paper. The problem is to design $u(t) \in
\Re$ and $v(t) \in \Re$ to effect some desired evolution. We now
show how to construct controls to give a rotation of angle
$\phi(\epsilon)$ around the $x$ axis or the $y$ axis of the Bloch
sphere. From these constructions, an arbitrary rotation on the Bloch
sphere can be constructed using an Euler angle decomposition.

\begin{figure}
\center
\includegraphics[scale=0.65]{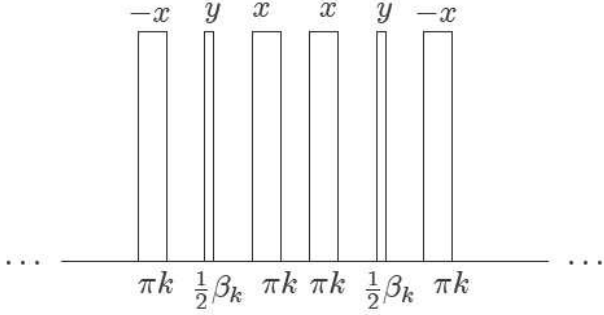}
\caption{A schematic depiction of the pulse sequence element in
(\ref{Uk_prop}). The pulse sequence element consists of six
individual rotations which can be produced using the controls $u(t)$
and $v(t)$ as explained in the text. \label{fig:eps_seq}}
\end{figure}

\subsection{Rotation About $y$ axis}
In a time interval $dt$, we can use the controls to generate
rotations $\exp(\epsilon u_0 dt \Omega_y)$ and $\exp(\epsilon v_0 dt
\Omega_x)$ where $u_0$ and $v_0$ are constants to be specified.
Using this idea, consider generating the rotation
\begin{equation}
U_k = U_{1k}U_{2k} \label{Uk_prop}
\end{equation}
with
\begin{eqnarray}
U_{1k} &=& \exp(-\pi k\epsilon \Omega_x) \exp(\frac{1}{2} \epsilon
\beta_k
\Omega_y) \exp(\pi k \epsilon \Omega_x) \label{u_1k_e} \\
U_{2k} &=& \exp(\pi k \epsilon \Omega_x) \exp(\frac{1}{2} \epsilon
\beta_k \Omega_y) \exp(-\pi k \epsilon \Omega_x) \label{u_2k_e}
\end{eqnarray}
using the controls $u$ and $v$.  Using the relation
\begin{eqnarray}
\exp(\alpha \Omega_x)\exp(\beta \Omega_y) \exp(-\alpha \Omega_x) = \nonumber \\
\exp(\beta(\cos (\alpha) \Omega_y + \sin(\alpha) \Omega_z ))
\end{eqnarray}
the matrices $U_{1k}$ and $U_{2k}$ may be rewritten as
\begin{eqnarray}
U_{1k} &=& \exp(\frac{1}{2} \epsilon \beta_k (\Omega_y \cos(\pi k
\epsilon) -
\Omega_z \sin(\pi k \epsilon))) \\
U_{2k} &=& \exp(\frac{1}{2} \epsilon \beta_k (\Omega_y \cos(\pi k
\epsilon) + \Omega_z \sin(\pi k \epsilon)))
\end{eqnarray}
For small $\beta_k$, we can make the approximation
\begin{equation}
U_{1k}U_{2k} \approx \exp(\epsilon \beta_k \cos(\pi k \epsilon)
\Omega_y) \label{approx}
\end{equation}
In (\ref{approx}), we have expanded the exponentials in $U_{1k}$ and
$U_{2k}$ to first order, performed the multiplication called for in
(\ref{Uk_prop}), and then rewritten the product as (\ref{approx})
keeping terms to first order. In the case when $\beta_k$ is too
large for (\ref{approx}) to represent a good approximation, we
should choose a threshold value $\beta_0$ such that (\ref{approx})
represents a good approximation and so that
\begin{equation}
\beta_k = n \beta_0
\end{equation}
with $n$ an integer.  Defining
\begin{eqnarray}
U_{10} &=& \exp(\frac{1}{2} \epsilon \beta_0 (\Omega_y \cos(\pi k
\epsilon) -
\Omega_z \sin(\pi k \epsilon))) \\
U_{20} &=& \exp(\frac{1}{2} \epsilon \beta_0 (\Omega_y \cos(\pi k
\epsilon) + \Omega_z \sin(\pi k \epsilon)))
\end{eqnarray}
we can apply the total propagator
\begin{eqnarray}
[ \ U_{10} U_{20} \ ]^n \label{approx_b01} \\
\approx \left[ \exp(\epsilon
\beta_0 \cos(\pi k \epsilon) \Omega_y)  \right]^n \label{approx_b0} \\
 = \exp(\epsilon \beta_k \cos(\pi k \epsilon) \Omega_y)
\end{eqnarray}
where we used the approximation (\ref{approx}) in (\ref{approx_b0}).
More will be said about this approximation below.

If we then think about making the incremental rotation $U_k$ for
many different values of $k$, we will get a net rotation
\begin{equation}
U = \prod_k \exp(\epsilon \beta_k \cos(\pi k \epsilon) \Omega_y)
\end{equation}
so long as we keep $\beta_k$ sufficiently small to justify the
approximation (\ref{approx}).  The total propagator $U$ for the
Bloch equations can then be rewritten as
\begin{equation}
U = \exp(\epsilon \sum_k \beta_k \cos(\pi k \epsilon) \Omega_y)
\label{propagator}
\end{equation}
If we now choose the coefficients $\beta_k$ so that
\begin{equation}
\sum_k \beta_k \cos(\pi k \epsilon) \approx
\frac{\phi(\epsilon)}{\epsilon}
\end{equation}
then we will have constructed a pulse sequence to approximate a
desired $\epsilon$ dependent rotation around the $y$ axis.  Since
$\epsilon$ is bounded away from the origin,
$\phi(\epsilon)/\epsilon$ is everywhere finite, and we can
approximate it with a Fourier Series.

\subsubsection{Remark About the Approximation}
Here we consider the error introduced by the approximation
(\ref{approx_b0}). Define the error $E(Z,V)$ when a unitary matrix
$V$ is implemented instead of a desired unitary matrix $Z$ by
\begin{equation}
E(Z,V) = \max_x \| (Z-V)x \|
\end{equation}
With these identifications, in (\ref{approx_b01}) we have
\begin{eqnarray}
V & = & U_{10}U_{20} \nonumber \\
 & = & I + \frac{\beta_k}{n} \cos(\pi k \epsilon) \Omega_y + M_1(n)
\end{eqnarray}
and
\begin{eqnarray}
Z & = & \exp(\beta_0 \cos(\pi k \epsilon) \Omega_y) \nonumber \\
 & = & I + \frac{\beta_k}{n} \cos(\pi k \epsilon) \Omega_y + M_2(n)
\end{eqnarray}
where $M_1(n)$ and $M_2(n)$ are matrices with finite entries and of
maximum order $\frac{1}{n^2}$. Notice this implies that the
difference $(Z-V)$ is of order $\frac{1}{n^2}$. As defined
previously, $\beta_k = n \beta_0$. A well known result (see for
example \cite{Ike}) says that the maximum error introduced by
implementing (multiplying) the product of $n$ of the $V$ matrices
instead of implementing $n$ of the $Z$ matrices is the sum of the
individual errors.  Thus, the total error $E_{\mathrm{total}}$
satisfies
\begin{eqnarray}
E_{\mathrm{total}} & \leq & n E(Z,V) \\
 & \sim & \frac{1}{n}
\end{eqnarray}
and thus, by making $n$ sufficiently large, we may decrease the
total error introduced in the above method to an arbitrarily small
value. For the simulations done in this paper, we find a value
$\beta_0 \leq 30 \deg$ produces good results.

\subsection{Rotation About $x$ axis}
An analogous derivation can be made for rotations about the $x$ axis
of the Bloch sphere.  Replacing (\ref{u_1k_e}) and (\ref{u_2k_e})
with
\begin{eqnarray}
U_{1k} &=& \exp(-\pi k\epsilon \Omega_y) \exp(\frac{1}{2} \epsilon
\beta_k \Omega_x)
\exp(\pi k \epsilon \Omega_y) \\
U_{2k} &=& \exp(\pi k \epsilon \Omega_y) \exp(\frac{1}{2} \epsilon
\beta_k \Omega_x) \exp(-\pi k \epsilon \Omega_y)
\end{eqnarray}
and following an analogous procedure leads to an approximate net
propagator
\begin{equation}
U = \exp(\epsilon \sum_k \beta_k \cos(\pi k \epsilon) \Omega_x)
\end{equation}
The coefficients $\beta_k$ may be chosen to approximate
$\phi(\epsilon) / \epsilon$, and thus we can approximately produce a
desired $\epsilon$ dependent rotation about the $x$ axis of the
Bloch sphere.  Since an arbitrary rotation on the Bloch sphere may
be decomposed in terms of Euler angles, the methods presented can be
used to approximately synthesize any evolution on the Bloch sphere.

\subsection{Choosing the Coefficients $\beta_k$}
Suppose we wish to design a pulse with a uniform net rotation angle
of $\phi$ around either the $x$ or $y$ axis of the Bloch sphere
using the previously discussed algorithm. We focus on the case of a
uniform rotation (independent of $\epsilon$) because this is the
most useful pulse sequence in NMR.  It is straightforward to
incorporate an $\epsilon$ dependent rotation $\phi(\epsilon)$ into
everything that follows. We face the problem of choosing $\beta_k$
and $k$ so that
\begin{equation}
\sum_k \beta_k \cos(\pi k \epsilon) \approx f(\epsilon), \ \
1-\delta \leq \epsilon \leq 1
\end{equation}
where
\begin{equation}
f(\epsilon) = \frac{\phi}{\epsilon}, \ \ 1-\delta \leq \epsilon \leq
1
\end{equation}
Since we only have $\cos(\pi k \epsilon)$ terms in the series, we
first will extend $f(\epsilon)$ to have even symmetry about
$\epsilon = 0$.  To do this, we define $g(\epsilon)$ to be
\begin{equation}
g(\epsilon) = \left\{ \begin{array} {r@{,\quad}l}
                            f(\epsilon) & 1-\delta \leq \epsilon \leq 1 \\
                            f(1-\delta) & -(1-\delta) \leq \epsilon \leq 1-\delta \\
                            f(-\epsilon) & -1 \leq \epsilon \leq -(1-\delta)
                            \end{array} \right.
                            \label{norelaxsol}
\end{equation}
and now consider choosing $\beta_k$ and $k$ so that
\begin{equation}
\sum_k \beta_k \cos(\pi k \epsilon) \approx g(\epsilon), \ \ -1 \leq
\epsilon \leq 1
\end{equation}
A natural choice is to choose $k$ as nonnegative integers, in which
case $\beta_k$ may be computed using the orthogonality relation
\begin{equation}
\int_{-1}^{1} \cos(\pi k \epsilon) \cos(\pi k^\prime \epsilon)
d\epsilon = \delta_{kk^\prime}, \ \ k \neq 0
\end{equation}
where $\delta_{k k^\prime}$ is the Kronecker delta. We find for the
coefficients
\[
\beta_k = \int_{-1}^{1} \cos(\pi k \epsilon) g(\epsilon) \ d
\epsilon , \ \ k \neq 0 \] \[ \beta_0 = \frac{1}{2} \int_{-1}^{1}
g(\epsilon) \ d \epsilon, \ \ k = 0
\]
The number of terms kept in the series is decided by the pulse
designer.
\begin{figure}
\center
\includegraphics[scale=0.5]{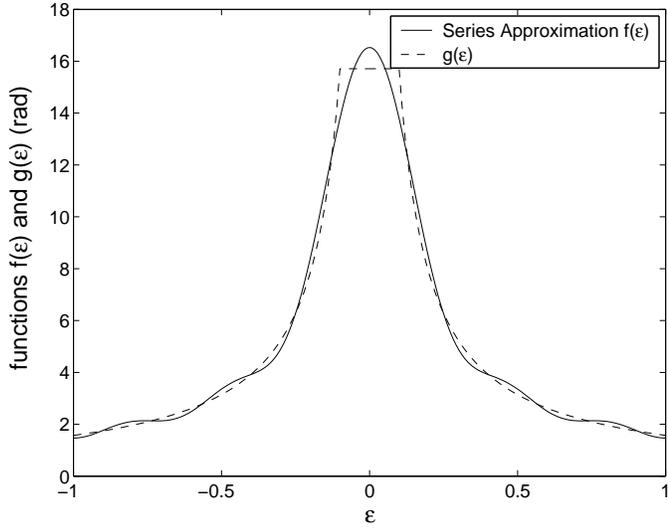}
\caption{An example design to approximate $g(\epsilon)$ over the
range $-1 \leq \epsilon \leq 1$ for the rotation angle $\phi = \pi /
2$. Five terms in the series expansion were retained. The relevant
range of rf inhomogeneity is $0.1 \leq \epsilon \leq 1$.
\label{fig:design}}
\end{figure}
A sufficient number of terms should be retained so that the error
across the relevant range of $\epsilon$ does not exceed some
acceptable value.  Figure \ref{fig:design} depicts an example design
using a series with five terms.  The region of interest for
$f(\epsilon)$ is $0.1 \leq \epsilon \leq 1$.  In this region, we see
relatively small errors. We now give two examples to demonstrate the
usefulness of the algorithm.

\subsection{Simulations}

\subsubsection{$\frac{\pi}{2}$ pulse around $y$ axis} 
\begin{figure}
\center
\includegraphics[scale=0.5]{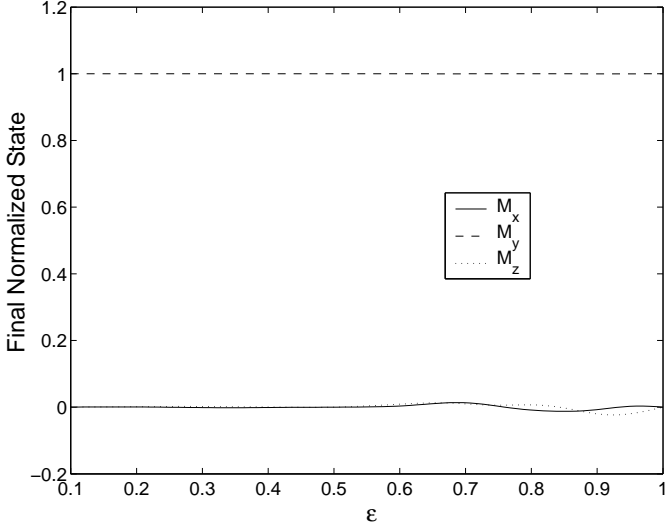}
\caption{Results of pulse sequence designed to produce uniform
$\pi/2$ rotation about $y$ axis. The sequence was applied to the
initial state $M(0) = [0 \ 0 \ 1]^T$, and the plot shows the final
state as a function of $\epsilon$ after propagating the Bloch
equations. \label{fig:design1}}
\end{figure}

\begin{figure}
\center
\includegraphics[scale=0.5]{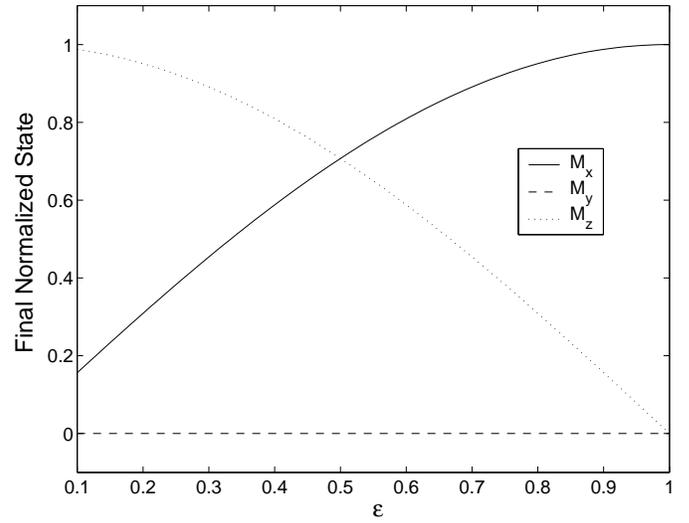}
\caption{Results of applying $u(t) = \frac{\pi}{2}$ for one unit of
time to the initial state $M(0) = [0 \ 0 \ 1]^T$.  The system
corresponding to the value $\epsilon = 1$ for the dispersion
parameter experiences a $\frac{\pi}{2}$ rotation about the $y$ axis
of the Bloch sphere, but systems corresponding to other values of
$\epsilon$ show deteriorated performance.
\label{fig:CDC_nieve_control_mod}}
\end{figure}

Suppose we wish
to design a $\frac{\pi}{2}$ pulse around the $y$ axis and we want to
consider rf inhomogeneity in the range $0.1 \leq \epsilon \leq 1$.
Then we should consider
\begin{equation}
f(\epsilon) = \frac{\pi}{2 \epsilon}, \ \ 0.1 \leq \epsilon \leq 1
\end{equation}
Figure \ref{fig:design1} shows the results of the designed pulse
sequence acting on the initial state $M(0) = [0 \ 0 \ 1]^T$ while
keeping five terms in the series expansion.
We see that the resulting pulse sequence reliably produces a net
evolution $\exp(\frac{\pi}{2}\Omega_y)$ across the entire range of
$\epsilon$ values.  Figure \ref{fig:CDC_nieve_control_mod} shows the
results of applying $u(t) = \frac{\pi}{2}$ for one unit of time to
the system (\ref{bloch_eps}).  This approach corresponds to assuming
the dispersion parameter $\epsilon$ is fixed at a nominal value
$\epsilon = 1$, so that every system sees the same control signals
$u(t)$ and $v(t)$.  Systems corresponding to values $\epsilon \neq
1$ exhibit deteriorated performance as demonstrated in Figure
\ref{fig:CDC_nieve_control_mod}.

\subsubsection{$\pi$ pulse} 
\begin{figure}
\center
\includegraphics[scale=0.5]{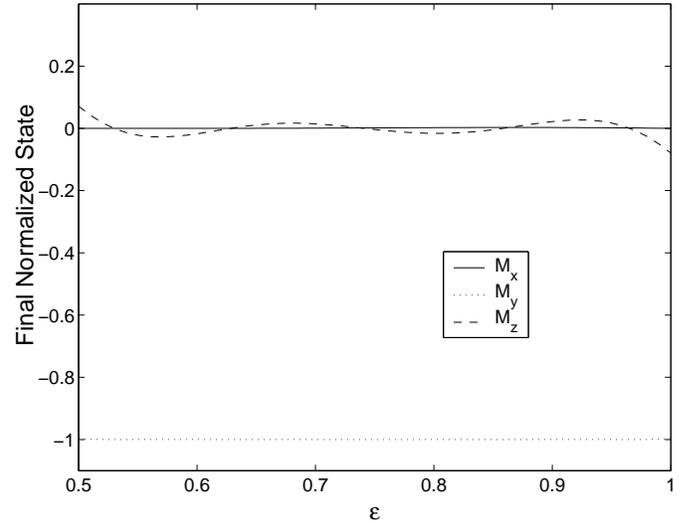}
\caption{Results of pulse sequence designed to produce uniform $\pi$
rotation about $x$ axis. The sequence was applied to the initial
state $M(0) = [0 \ 1 \ 0]^T$, and the plot shows the final state as
a function of $\epsilon$ after propagating the Bloch equations.
\label{fig:design4}}
\end{figure}

As a second example, suppose we wish
to design a $\pi$ pulse around the $x$ axis and we want to consider
rf inhomogeneity in the range $0.5 \leq \epsilon \leq 1$. Then we
should consider
\begin{equation}
f(\epsilon) = \frac{\pi}{\epsilon}, \ \ 0.5 \leq \epsilon \leq 1
\end{equation}
Figure \ref{fig:design4} shows the results of the designed pulse
sequence acting on the initial state $M(0) = [0 \ 1 \ 0]^T$ while
keeping nine terms in the series expansion.
We see that the resulting pulse sequence reliably produces a net
evolution $\exp(\pi \Omega_y)$ across the entire range of $\epsilon$
values.

It should be noted that although we consider design examples where
we wish to produce a uniform rotation that is independent of the
parameter $\epsilon$, the method presented in the paper can also be
used to design control laws which prepare the final state as a
function of the parameter $\epsilon$.  We consider examples to
produce a uniform rotation, independent of $\epsilon$, because this
is the most useful application in NMR.

\section{Design Method for Position Dependent Rotations}
Now consider the Bloch equations with no rf inhomogeneity and with a
linear gradient
\begin{equation}
\frac{d}{dt}\left[\begin{array}{c} M_x  \\
M_y \\
M_z \end{array}\right] = (G(t) s \Omega_z + u(t) \Omega_y + v(t) \Omega_x)  \left[\begin{array}{c} M_x \\
M_y \\
M_z \end{array}\right]
\end{equation}
where $u(t) \in \Re$, $v(t) \in \Re$, and $G(t) \in \Re$ are time
dependent control amplitudes we may specify, and $s \in [0, 1]$ can
be thought of as a dispersion parameter.  As previously discussed,
$s$ represents the spatial position of the spin system in the sample
under study. The goal is to engineer a control law that will effect
a net position dependent rotation, so that the final state is
prepared as a function of $s$.  A common example in NMR imaging is a
so-called slice selective sequence, whereby the controls should
selectively perform a $\frac{\pi}{2}$ rotation on some range of $s$
values, while performing no net rotation on $s$ values falling
outside of that range.

Using the controls, consider generating the evolution
\begin{equation}
U_k = U_{1k}U_{2k}
\end{equation}
with
\begin{eqnarray}
U_{1k} &=& \exp(\pi ks \Omega_z) \exp(\frac{1}{2}\beta_k \Omega_y)
\exp(-\pi ks \Omega_z) \label{u_1k_s} \\
U_{2k} &=& \exp(-\pi ks \Omega_z) \exp(\frac{1}{2} \beta_k \Omega_y)
\exp(\pi ks \Omega_z) \label{u_2k_s}
\end{eqnarray}
The matrices $U_{1k}$ and $U_{2k}$ may be rewritten as
\begin{eqnarray}
U_{1k} &=& \exp(\frac{1}{2} \beta_k(\cos(\pi ks) \Omega_y - \sin(\pi ks) \Omega_x)) \\
U_{2k} &=& \exp(\frac{1}{2} \beta_k(\cos(\pi ks) \Omega_y + \sin(\pi
ks) \Omega_x))
\end{eqnarray}
Again performing a first order analysis on the exponentials as was
done in the previous section, we can make the approximation
\begin{equation}
U_{1k}U_{2k} \approx \exp(\beta_k \cos(\pi ks) \Omega_y)
\end{equation}
If we then think about making the rotation $U_k$ for different
values of $k$ and $\beta_k$, we will get the net propagator
\begin{equation}
U = \prod_k \exp(\beta_k \cos(\pi k s) \Omega_y) \label{prop_x_y}
\end{equation}
within the approximation previously discussed. The propagator
(\ref{prop_x_y}) may be rewritten as
\begin{equation}
U = \exp(\sum_k \beta_k \cos(\pi ks) \Omega_y)
\end{equation}
Choosing $k$ as the nonnegative integers, and choosing the $\beta_k$
so that
\begin{equation}
\sum_k \beta_k \cos(\pi ks) \approx \phi(s)
\end{equation}
where $\phi(s)$ is the desired position $s$ dependent rotation angle
results in a net rotation around the $y$ axis of the Bloch sphere
with the desired dependence on the parameter $s$.

An analogous procedure may be used to generate an $s$ dependent
rotation around the $x$ axis of the Bloch sphere.  Replacing
(\ref{u_1k_s}) and (\ref{u_2k_s}) with
\begin{eqnarray}
U_{1k} &=& \exp(-\pi ks \Omega_z) \exp(\frac{1}{2} \beta_k \Omega_x)
\exp(\pi ks
\Omega_z) \\
U_{2k} &=& \exp(\pi ks \Omega_z) \exp(\frac{1}{2} \beta_k \Omega_x)
\exp(-\pi ks \Omega_z)
\end{eqnarray}
and following a similar procedure, we can approximately produce the
total propagator
\begin{equation}
U = \exp(\sum_k \beta_k \cos(\pi ks) \Omega_x)
\end{equation}
Choosing $k$ as the nonnegative integers, and choosing the $\beta_k$
appropriately results in a net rotation around the $x$ axis of the
Bloch sphere with the desired dependence on the parameter $s$. Since
any rotation can be decomposed in terms of Euler angles, we may use
the methods just discussed to approximately produce any position $s$
dependent rotation on the Bloch sphere.

\subsection{Design Example}
\begin{figure}
\center
\includegraphics[scale=0.5]{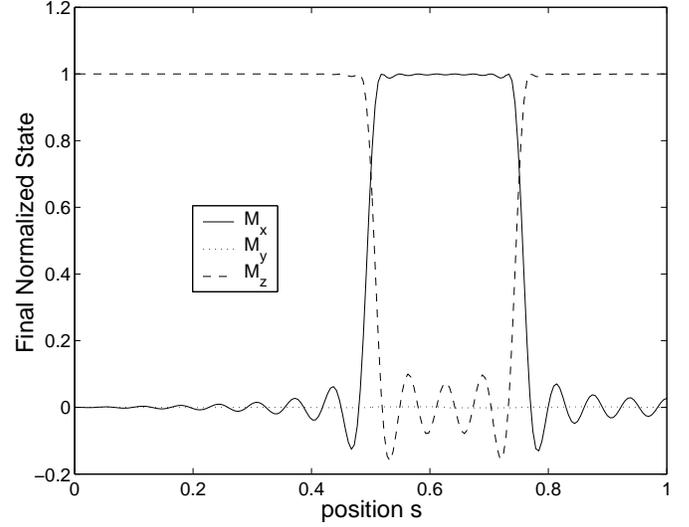}
\caption{Results of pulse sequence designed to produce uniform
$\frac{\pi}{2}$ rotation about $y$ axis over the range $0.5 \leq s
\leq 0.75$. The sequence was applied to the initial state $M(0) = [0
\ 0 \ 1]^T$, and the plot shows the final state as a function of $s$
after propagating the Bloch equations. \label{fig:xfigure}}
\end{figure}

As an example design using the procedure just discussed, consider a
slice selective pulse sequence, where we wish to excite a certain
range of $s$ values while leaving systems with $s$ values falling
outside of that range unaffected at the end of the sequence.
\begin{equation}
\phi(s) = \left\{ \begin{array} {r@{,\quad}l}
                            \frac{\pi}{2} & 0.5 \leq s \leq 0.75 \\
                            0 & \mathrm{otherwise}
                            \end{array} \right.
                            \label{norelaxsol}
\end{equation}
Figure \ref{fig:xfigure} shows the results of a pulse sequence
designed using the procedure described in the text while keeping 30
terms in the series.
The ripples appearing in Figure \ref{fig:xfigure} result from the
ripples in the approximation of the sharp slice selective profile
using a Fourier Series.  One method used to overcome this in
practice is to allow for a ramp between the $0$ and $\frac{\pi}{2}$
level on the slice.

\section{Control Laws Involving Position and rf Inhomogeneity}
We now come back to the problem of considering the full version of
the Bloch equations (\ref{chap32_bloch}) including the two
dispersion parameters $s$ and $\epsilon$. Rewriting
(\ref{chap32_bloch}) in terms of the generators of rotation we have

\begin{equation}
\frac{d}{dt}\left[\begin{array}{c} M_x  \\
M_y \\
M_z \end{array}\right] = (G(t) s \Omega_z + \epsilon u(t) \Omega_y + \epsilon v(t) \Omega_x)  \left[\begin{array}{c} M_x \\
M_y \\
M_z \end{array}\right] \label{bloch_xe}
\end{equation}
where the state vector is now a function of both parameters $s$ and
$\epsilon$.  The control task is to choose $u(t) \in \Re$, $v(t) \in
\Re$, and $G(t) \in \Re$ to effect a desired rotation
$\phi(s,\epsilon)$.  Proceeding along the lines of the previous two
sections, consider generating the propagators
\begin{eqnarray}
U_{1k} &=& \exp(\pi k_1 s \Omega_z) \exp(\frac{1}{4} \epsilon
\beta_k
\Omega_y) \exp(-\pi k_1 s \Omega_z) \nonumber \\
 &=& \exp(\frac{1}{4} \epsilon \beta_k (\cos(\pi k_1 s)\Omega_y - \sin(\pi k_1 s)\Omega_x))
\end{eqnarray}
and
\begin{eqnarray}
U_{2k} &=& \exp(-\pi k_1 s \Omega_z) \exp(\frac{1}{4} \epsilon
\beta_k
\Omega_y)\exp(\pi k_1 s \Omega_z) \nonumber \\
 &=& \exp(\frac{1}{4} \epsilon \beta_k (\cos(\pi k_1 s)\Omega_y + \sin(\pi k_1 s)\Omega_x))
\end{eqnarray}
Within a first order approximation for the exponentials, we have the
approximate total propagator
\begin{equation}
U_{1k}U_{2k} \approx \exp(\frac{1}{2} \epsilon \beta_k \cos(\pi k_1
s) \Omega_y)
\end{equation}
Building on this, we can produce the propagator
\begin{eqnarray}
U_{3k} &=& \exp(\pi k_2 \epsilon \Omega_x) U_{1k}U_{2k} \exp(-\pi
k_2 \epsilon
\Omega_x) \nonumber \\
 & \approx & \exp(\frac{1}{2} \epsilon \beta_k \cos(\pi k_1 s) (\cos(\pi k_2 \epsilon) \Omega_y - \sin(\pi k_2 \epsilon)
 \Omega_z)) \nonumber
\end{eqnarray}
Similarly, we can produce
\begin{eqnarray}
U_{4k} &=& \exp(-\pi k_2 \epsilon \Omega_x) U_{1k}U_{2k} \exp(\pi
k_2 \epsilon
\Omega_x) \nonumber \\
 & \approx & \exp(\frac{1}{2} \epsilon \beta_k \cos(\pi k_1 s) (\cos(\pi k_2 \epsilon) \Omega_y + \sin(\pi k_2 \epsilon)
 \Omega_z)) \nonumber
\end{eqnarray}
so that we can approximately produce the total propagator
\begin{eqnarray}
U_k &=& U_{3k}U_{4k} \nonumber \\
 & \approx & \exp(\epsilon \beta_k
\cos(\pi k_1 s) \cos(\pi k_2 \epsilon) \Omega_y)
\end{eqnarray}
within the approximation for the exponentials.  We can use the
method previously discussed in the case when $\beta_k$ is too large
for the approximation to be valid. Producing the propagator $U_k$
for different values of $k_1$, $k_2$, and $\beta_k$ results in the
net propagator
\begin{equation}
U = \prod_{\{k_1,k_2\}} \exp(\epsilon \beta_k \cos(\pi k_1 s)
\cos(\pi k_2 \epsilon) \Omega_y)
\end{equation}
A choice of $k_1$, $k_2$, and $\beta_k$ so that
\begin{equation}
\sum_{ \{k_1,k_2 \}} \beta_k \cos(\pi k_1 s) \cos(\pi k_2 \epsilon)
\approx \frac{\phi(s,\epsilon)}{\epsilon}
\end{equation}
where $\phi(s,\epsilon)$ is the desired position $s$ and rf
inhomogeneity parameter $\epsilon$ dependent rotation angle, results
in an approximate desired evolution for the Bloch equations
(\ref{bloch_xe}).

Analogous arguments show we may approximately produce a rotation
around the $x$ axis of the Bloch sphere
\begin{equation}
U_k = \exp(\epsilon \beta_k \cos(\pi k_1 s) \cos(\pi k_2
\epsilon)\Omega_x)
\end{equation}
and may thus approximately generate a net propagator
\begin{equation}
U = \prod_{ \{k_1,k_2 \}} \exp(\epsilon \beta_k \cos(\pi k_1 s)
\cos(\pi k_2 \epsilon) \Omega_x)
\end{equation}
and thus approximately produce a desired position $s$ and rf
inhomogeneity parameter $\epsilon$ rotation around the $x$ axis of
the Bloch sphere.  Since any rotation on the unit sphere can be
decomposed in terms of Euler angles, an arbitrary $(s, \epsilon)$
dependent rotation can be approximately produced using these
methods.

\section{Conclusions}
In this paper we have provided new methods to design control laws
for the Bloch equations when certain dispersion parameters are
present in the system dynamics.  These methods are of utmost
practical importance in the fields of NMR spectroscopy and NMR
imaging, and can be implemented in many well known experiments
immediately.  The methods presented in the paper allow the design of
a compensating control law (pulse sequence) that will compensate for
dispersions in the system dynamics while providing a clear tradeoff
for the control law designer between total time required for the
sequence and amplitude of the available controls.


\end{document}